# From Anomalous to Normal: Temperature Dependence of the Band Gap in Two-Dimensional Black Phosphorus

Shenyang Huang[1], Fanjie Wang[1], Guowei Zhang[2,1], Chaoyu Song[1], Yuchen Lei[1], Qiaoxia Xing[1], Chong Wang[1], Yujun Zhang[3], Jiasheng Zhang[1], Yuangang Xie[1], Lei Mu[1], Chunxiao Cong[4], Mingyuan Huang[3]* & Hugen Yan[1]*

[1]State Key Laboratory of Surface Physics, Key Laboratory of Micro and Nano - Photonic Structures (Ministry of Education), and Department of Physics, Fudan University, Shanghai 200433, China

[2]Institute of Flexible Electronics, Northwestern Polytechnical University, Xi'an 710072, China

[3]Department of Physics, Southern University of Science and Technology, Shenzhen 518055, China

[4]State Key Laboratory of ASIC and System, School of Information Science and Technology, Fudan University, Shanghai 200433, China

*E-mail: hgyan@fudan.edu.cn (H. Y.), huangmy@sustech.edu.cn (M. H.)

The temperature dependence of the band gap is crucial to a semiconductor. Bulk black phosphorus (BP) is known to exhibit an anomalous behavior. Through optical spectroscopy, here we show that the temperature effect on BP band gap gradually evolves with decreasing layer number, eventually turns into a normal one in the monolayer limit, rendering a crossover from the anomalous to the normal. Meanwhile, the temperature-induced shift in optical resonance also differs with different transition indices for the same thickness sample. A comprehensive analysis reveals that the temperature-tunable interlayer coupling is responsible for the observed diverse scenario. Our study provides a key to the apprehension of the anomalous temperature behavior in certain layered semiconductors.

Black phosphorus (BP) is a layered material with intriguing properties, such as remarkable in-plane anisotropy and strongly layer-dependent band structure[1,2]. Different from 2-dimensional transition metal dichalcogenides (TMDC), BP is always a direct gap semiconductor regardless of the thickness. Interestingly, there are a series of optical resonances in few-layer BP due to the band splitting induced by the interlayer coupling[3-5], which offers us an unambiguous signature to monitor the external perturbations to such coupling[6].

Temperature is a tuning knob to the band gap, which plays crucial roles to the performance of semiconductor devices. In most cases, the band gap decreases with increasing temperature[7-9]. However, anomalous behavior appears in certain semiconductors, whose band gap increases with temperature[10,11]. Among them is bulk BP[12-15]. Thermal expansion is believed to play a decisive role[14]. This scenario seems consistent with strain experiments for few-layer BP, which shows in-plane tensile strain enlarges the band gap[6,16]. Along this line, it's plausible to believe that monolayer and few-layer BP would resemble bulk BP in the temperature dependence of the band gap.

In this study, we systematically investigated the temperature dependence of the band structure in monolayer, few-layer and bulk BP on quartz substrates through optical spectroscopy. To our surprise, monolayer and few-layer BP (layer number < 5) behave qualitatively differently from the bulk. The band gap of monolayer BP decreases monotonically with increasing temperature, exhibiting a normal temperature dependence. By monitoring the peak separation of few-layer BP as a function of temperature, we can attribute the temperature-tunable interlayer coupling to the anomaly in thick BP samples. Moreover, special care has been taken to account for the extrinsic strain effect induced by the substrate, for which, multiple substrates have been compared. Our comprehensive study not only sheds light on the mechanisms of anomalous temperature behavior, but also paves a way for versatile vdWs engineering of few-layer BP through thermal effects.

Figs. 1(a) and 1(b) show the optical spectra of monolayer and bulk BP under various temperatures, ranging from 10 to 300 K (for details about measurement refer to method in Supplementary Information [17]). In monolayer BP, the prominent optical resonance due to the exciton (labeled as $E_{11}$) can be clearly distinguished, and the single particle bandgap has a higher energy than that with a separation of the exciton binding energy[23,24]. Since the exciton binding energy has little dependence on temperature[24,25], the change of the exciton peak position can be fully attributed to the band gap effect. This is also applicable to few-layer BP as shown in Fig. 2. Therefore, for simplicity we directly treat the optical resonances as the onset of band-to-band transitions without further differentiation. Fig. 1(c) shows the bandgap of mono- and bulk BP as a function of temperature (the band gap of monolayer can be extracted through Lorentzian fitting and the extraction of the gap for the bulk is detailed in Supplementary Information [17]). As we can see, the band gap of monolayer BP shifts monotonically to the blue with decreasing temperature, which is in sharp contrast to bulk BP[12-15]. Very interestingly, for the same material, different thickness renders qualitatively different temperature dependence.

Indeed, the thickness dependence of the temperature effect is systematic. In addition to monolayer and bulk BP, we also studied few-layer BP on quartz substrates with layer number up to 10. Fig. 2(c) shows the infrared extinction spectra of the BP flake shown in Fig. 2(a), where optical transitions associated with different layer number and subband indices can be clearly distinguished (details in Supplementary Information [17]). Due to the interlayer coupling, the valence (conduction) band splits into multiple well-separated subbands in few-layer BP. As a result, besides the $E_{11}$, the infrared spectra show multiple resonance peaks, denoted by $E_{22}$…, schematically illustrated in Fig. 2(b). We also present all optical resonance positions as a function of temperature in Fig. 2(d). As we can see, the temperature dependence of the band gap exhibits strong layer-dependence. The overall trend in bilayer BP is similar to that in monolayer BP, i.e., $E_{11}$ increases with decreasing temperature. While for 4L BP, the band gap evolves non-monotonically with temperature, with a maximum at ~140 K. For 5L BP, the temperature dependence is nonmonotonic as well, but the overall trend gets closer to that of the bulk. Besides, we see that the trend of $E_{22}$ in 5L BP is totally different from that of $E_{11}$, but almost the same as $E_{11}$ of 2L BP. This underlines that the temperature effect also depends on subband indices, which is further verified in a 9L case shown in Fig. 2(e), where three optical transitions, namely $E_{11}$, $E_{22}$ and $E_{33}$, shift differently with temperature. In fact, such layer- and transition index-dependent temperature effect is true for all measured samples (more than 40) with different thickness (see Fig. S4).

Interestingly, although the temperature dependence looks so diverse for different thickness and transition index, the energy difference between optical transitions shows a linear relation with temperature. As shown in Figs. 3(b) and 3(c), the energy differences of $E_{11}$ between 2L and 4L, 2L and 5L and $E_{22}$-$E_{11}$ in 5L, 9L, $E_{33}$-$E_{11}$ in 9L, all change linearly with temperature. The energy difference between two peaks decreases with increasing temperature. In addition, the slope of the linearly fitted lines increases with the energy difference of the two transitions. We carefully examine the relationship between the energy difference and temperature and it always shows an excellent linear trend. The extracted slopes of the linear fittings, which describe the changing rates of the energy difference versus temperature, are plotted in Fig. 3(d) as a function of the energy difference between those two optical transitions at room temperature. All data points, including those from other thickness samples (<11L), fall on a straight line (red line in Fig. 3(d)) with a slope of -2.4×10$^{-4}$/K and zero intercept.

Next, we will examine the mechanisms responsible for the scenario. As we mentioned earlier, the energy difference between different optical resonances in the same few-layer BP is related to the interlayer interaction, or more quantitatively, proportional to the interlayer hopping parameter $t_\perp$ (Ref.[5]). As a result, the peak separation becomes an unambiguous indicator of the interlayer coupling. As Figs. 3(b) and 3(c) show (also Fig. S4), the energy spacings of different optical transitions change linearly with temperature, indicating the linearly temperature-dependent interlayer interaction. This is reasonable, given that the interlayer distance changes only slightly and linearly with temperature in our experimental range due to thermal expansion[26,27]. More specifically, when the temperature increases, BP will expand along the out-of-plane direction, increasing the interlayer distance and hence weakening the interlayer coupling (Fig. 3(a)). Indeed, such weakening of interlayer coupling has been observed in other 2D materials as well, though with quite different experimental manifestations[28,29].

To have a quantitative analysis, we use the tight-binding model to account for the interlayer coupling, which has successfully described the band structure evolution of few-layer BP[3,4,6]. According to this model, the optical transition energy at $\Gamma$ point of the Brillouin zone reads:

$$E_{nn}^{N}(T) = E_{g0}(T) - \Delta\gamma(T)\cos(\frac{n\pi}{N+1}) \qquad (1)$$

where $E_{g0}(T)$ is the bandgap of monolayer BP at temperature $T$, $\Delta\gamma(T)$ is the difference of overlapping integrals between conduction band ($\gamma_c$) and valence band ($\gamma_v$), which is proportional to the interlayer hopping parameter ($t_\perp$), $n$ is the transition (subband) index and $E_{nn}^{N}$ denotes the optical transition from the $n$th valence subband to the $n$th conduction subband for a $N$-layer BP. Based on our previous study[4], the fitted values of $E_{g0}$ and $\Delta\gamma$ at room temperature are 2.12 eV and 1.76 eV, respectively. According to previous studies [29,30], the thermal expansion of few-layer TMDC is comparable to their bulk counterparts, hence we implicitly assume in equation (1) that the lattice constants of each thickness sample are the same for BP at temperature $T$, independent of the thickness. Since our data indicate that the interlayer hopping changes linearly with temperature, we can rewrite $\Delta\gamma(T)$ as $\Delta\gamma(T) = \Delta\gamma_0 + hT$, where $h$ is the changing rate of $\Delta\gamma$ ($h < 0$). By substituting $\Delta\gamma(T)$ into Eq. (1), we get:

$$E_{nn}^{N} = E_{g0}(T) - h\cos(\frac{n\pi}{N+1})T + C \qquad (2)$$

where $C = -\Delta\gamma_0 \cos(\frac{n\pi}{N+1})$, which is independent of $T$. Now the temperature effect can be decomposed into two parts, one is related to the monolayer bandgap $E_{g0}(T)$, which has no layer- and transition-index dependence and is present for each transition. The other is a linear term $-h\cos(\frac{n\rho}{N+1})T$ with an $n$- and $N$-dependent slope of $-h\cos(\frac{n\rho}{N+1})$. Clearly, the layer- and transition index-dependent temperature effect results from the non-zero coefficient $h$. For the band gap of bulk BP, the linear term is $-hT$ ($n = 1$ and $N = \infty$), which increases with temperature (note that $h < 0$). This is the origin of the anomalous temperature dependence for bulk BP. Previously, the volumetric effect is usually phenomenologically attributed to the anomalous temperature dependence[11]. Here we precisely pin down the mechanism for BP, which provides a fresh perspective on the anomaly. Moreover, it also gives us some interesting insights. For instance, we should have the same temperature dependence for optical resonances when $n/(N+1)$ is the same but with different ($n$, $N$) pair, e.g., 2L $E_{11}$ and 5L $E_{22}$. Indeed, this is what we exactly observed in Fig. 2(d). The same scenario can be also found in other ($n$, $N$) pairs, such as (3L $E_{11}$, 7L $E_{22}$) and (4L $E_{11}$, 9L $E_{22}$) (see Fig. S5). From Eqs. (2) and (1), we have

$$\frac{d(E_{nn}^N - E_{mm}^M)}{dT} = A(E_{nn}^N - E_{mm}^M)|_{T=300K} \quad (3)$$

where $A = \frac{h}{\Delta g_{T=300K}}$. $\Delta\gamma_{T=300K}$ and $(E_{nn}^N - E_{mm}^M)|_{T=300K}$ are the values of $\Delta\gamma$ and the energy difference at 300 K, respectively. By linearly fitting the data points in Fig. 3(d), we can obtain $A$ of -2.4 x $10^{-4}$ /K, corresponding to -0.4 meV/K for the parameter $h$. When the temperature decreases from 300 to 10 K, $\Delta\gamma$ increases by 116 meV, amounting to a remarkable 6% change of the original value. Once $h$ is obtained, the temperature effect due to the interlayer coupling is fully accounted. Therefore, we can easily get the intralayer contribution of the temperature dependence $E_{g0}(T)$ by subtracting the interlayer ones in Eq. (2), which should be consistent with the monolayer BP case shown in Fig. 1(c). Indeed, this is verified in Fig. 3(e), which shows that all the intralayer components obtained from different transitions of samples with a variety of thickness coincide.

Now let us take a closer look at the intralayer component $E_{g0}(T)$. For the temperature dependence of monolayer BP, there should be two parts. One reflects the electron-phonon interaction, which shrinks the band gap when the temperature increases and is responsible for the normal behavior in most semiconductors[8]. The other accounts for the thermal expansion. Given the tiny thermal expansion coefficients of quartz substrate (~0.5 x $10^{-6}$ /K)[31], the in-plane thermal expansion of BP could be neglected, since BP layers are believed to be in good contact with the substrate[6]. Instead, the out-of-plane thermal expansion plays a role due to the puckered structure. The monolayer thickness $d$ (illustrated in Fig. 3(a)), which denotes the distance between the two sublayers, can change with temperature. Recently, a study revealed that the BP surface buckling can also affect the bandgap[32]. According to their results, the buckling expands with increasing temperature and enlarges the monolayer bandgap. While in our case, the monolayer bandgap shrinks with increasing temperature, indicating the dominant roles are played by electron-phonon interaction and thermal expansion.

We use the formula $E_{g0}(T) = \dfrac{aq}{e^{q/T}-1} + mT + c$ to fit the monolayer band gap, in which we choose the one-oscillator model[8] to describe the electron-phonon interaction and a linear term to account for the thermal expansion of BP[26,33]. Parameter $c$ is the monolayer band gap at zero temperature. Since we mainly focus on the first two terms, the last parameter $c$ can be eliminated by subtracting $E_{g0}(300K)$, as has been done in Fig. 3(e). Through a global fitting of the intralayer temperature dependence obtained from mono-, few-layer and bulk BP in Fig. 3(e), we obtain the fitting parameters $\alpha$, $\Theta$ and $m$ of -0.32 meV/K, 684 K and -0.18 meV/K, respectively. It is worth noting that $\Theta$ is the effective frequency for the dominant phonon mode[34], whose value (684 K ~456 cm$^{-1}$) is quite close to $A_g^2$ Raman mode (frequency ~468 cm$^{-1}$)[35,36]. This suggests that the $A_g^2$ mode has a dominant contribution to the electron-phonon interaction, which is fully consistent with previous work[36,37]. In addition, parameter $m$ is also very reasonable, as detailed in the Supplementary Information [17].

With those obtained parameters, we can reproduce the temperature dependence of all optical transitions from monolayer to bulk BP and the calculated curves show excellent agreement with our experimental data (see Fig. S6). It clearly shows that the temperature dependence of $E_{11}$ (band gap) evolves from normal one to anomalous one when the thickness increases from monolayer to bulk, and the parameter $h$ is fully responsible for such scenario.

Finally, we will examine the temperature dependence of the band structure in few-layer BP on different substrates. In addition to quartz, sapphire and $BaF_2$ with significantly larger thermal expansion coefficients are also employed for comparisons (see Supplementary Figs. S7 and S8 [17]). Interestingly, few-layer BP shows distinctive temperature dependence on different substrates, which is exemplified by the $E_{11}$ transition in 7L BP samples on those substrates, as shown in Fig. 4a (for more data refer to Fig. S10). For the latter two substrates, additional in-plane strain is induced when the temperature changes. Our previous study shows the in-plane strain not only modifies the intralayer bonding, but also the interlayer coupling[6] as shown in Fig.4(b), which complicates the pure temperature effect and leads to the quantitatively different behavior. More data are presented in Fig. S9 in Supplementary Information [17] with related discussions.

In summary, we systematically investigated the temperature dependence of the band structure from bulk to monolayer BP. We find that the interlayer interaction is sensitive to temperature, which induces strong layer- and transition index- dependence for the temperature effect on the band structure. Interestingly, the anomalous temperature dependence for the bulk and relatively thick BP layers can be fully attributed to the temperature tunable interlayer coupling. Our results shed light on the importance of vdWs coupling in defining the electronic structures of 2D materials and pave the way to versatile vdWs engineering.

## Acknowledgments

H.Y. is grateful to the financial support from the National Natural Science Foundation of China (Grant Nos. 11874009, 11734007), the National Key Research and Development Program of China (Grant Nos. 2016YFA0203900 and 2017YFA0303504), Strategic Priority Research Program of Chinese Academy of Sciences (XDB30000000), and the Oriental Scholar Program from Shanghai Municipal Education Commission. C. C. acknowledges the financial support from the Shanghai Municipal Science and Technology Commission (Grant No. 18JC1410300), the National Natural Science Foundation of China (Grant No. 61774040), the Fudan University-CIOMP Joint Fund (Grant No. FC2018-002), and the National Young 1000 Talent Plan of China. G.Z. acknowledges the financial support from the National Natural Science Foundation of China (Grant No. 11804398), Natural Science Basic Research Program of Shaanxi (Grant No. 2020JQ-105), Open Research Fund of State Key Laboratory of Surface Physics and the Fundamental Research Funds for the Central Universities. C.W. is grateful to the financial support from the National Natural Science Foundation of China (Grant No. 11704075) and China Postdoctoral Science Foundation. Part of the experimental work was carried out in Fudan Nanofabrication Lab.

# Figures

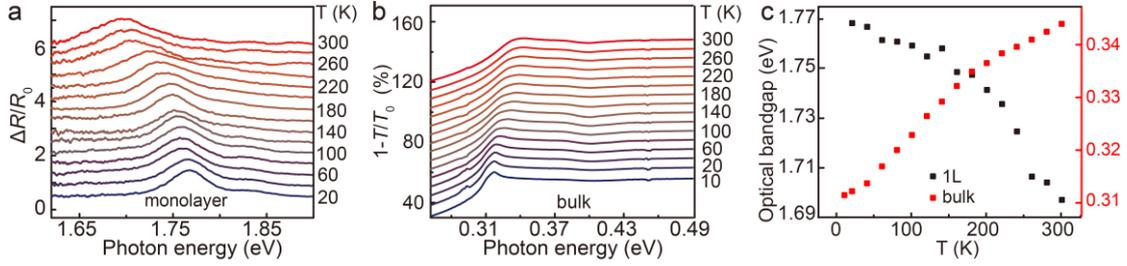

**Fig. 1. Temperature dependence of bandgaps in monolayer and bulk BP.** (a) Reflection spectra of monolayer BP and (b) extinction spectra of bulk BP (thickness ~40 nm) under different temperatures. Step is 20K from 20K to 300K. For clarity, the spectra are offset vertically. (c) Optical bandgaps of the monolayer and the bulk as a function of temperature.

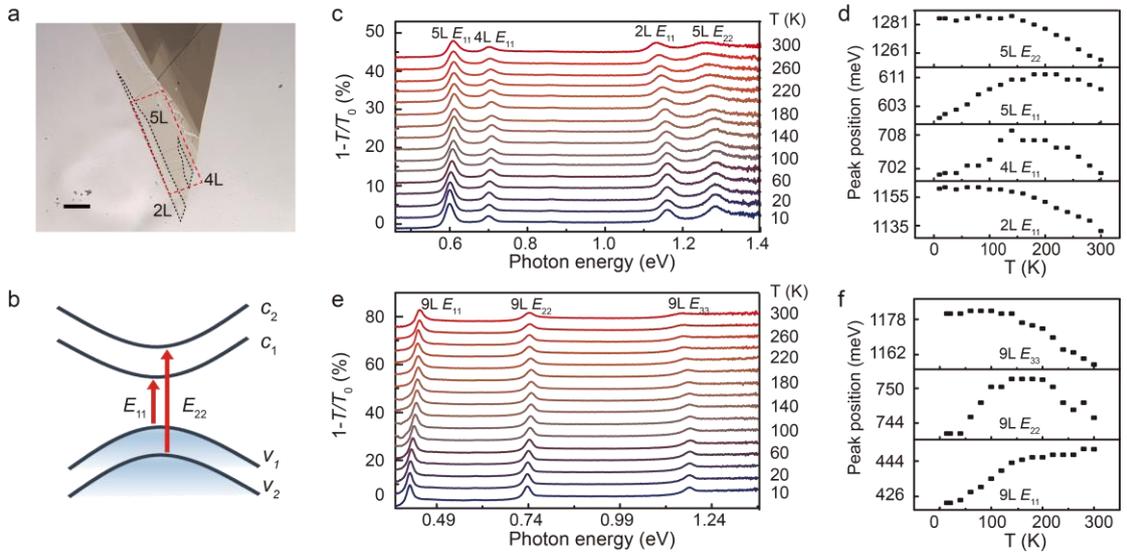

**Fig. 2. Layer- and transition index-dependent temperature effect in few-layer BP.** (a) Optical image of a BP sample on quartz substrate. The BP flake contains connected 2L, 4L and 5L parts in the same region. Scale bar is 10 μm. The region inside the red box is where the IR light shines. (b) Illustration of optical transitions between subbands in few-layer BP. (c) Infrared extinction spectra of the sample in Fig. 2(a) under different temperatures. Step is 20K from 20K to 300K. For clarity, the spectra are offset vertically. (d) Peak positions versus temperature for $E_{11}$ of 2L, 4L, 5L and $E_{22}$ of 5L. (e) Infrared

extinction spectra of a 9L BP under different temperatures. (f) Peak positions of $E_{11}$, $E_{22}$ and $E_{33}$ of 9L BP versus temperature.

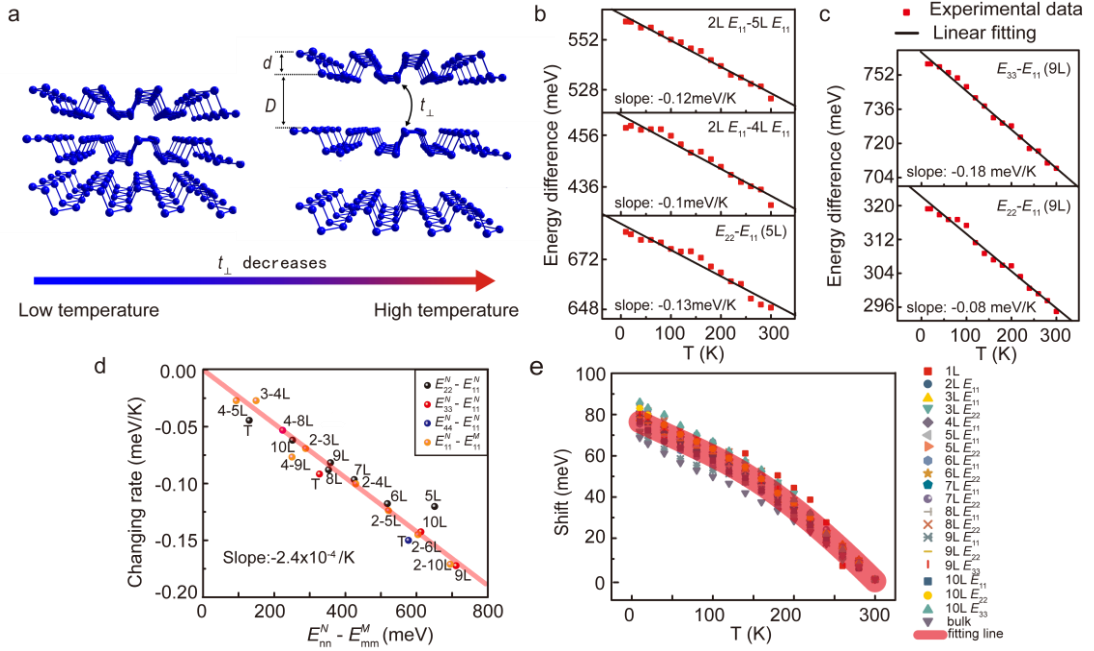

**Fig. 3. Interlayer and intralayer contributions to the temperature effect on the band structures.** (a) Cartoon illustration for the evolution of the atomic structure of trilayer BP with temperature. (b), (c) Energy difference of different optical transitions shown in Fig. 2 versus temperature. (d) Linear changing rate of peak energy difference (extracted from the linear fittings of $E_{nn}^{N} - E_{mm}^{M}$ versus temperature) as a function of the energy difference ($E_{nn}^{N} - E_{mm}^{M}$). In the figure, data point 'T' is a relatively thick BP whose layer number is ~20. (e) Intralayer contribution to the temperature effect obtained from the monolayer and other thickness BP samples. Y-axial represents the monolayer peak position shift with respect to 300 K, i.e., $E_{g0}(T)$-$E_{g0}$(300K).

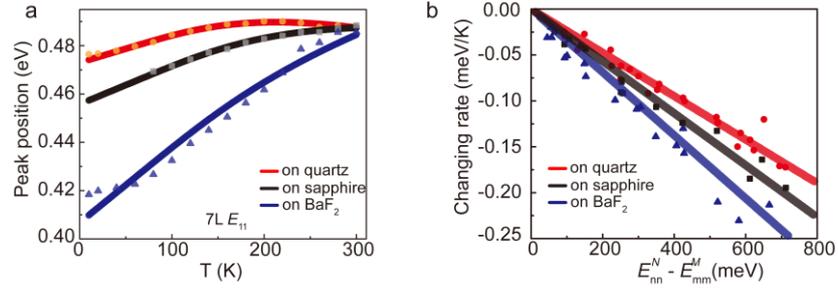

**Fig. 4. Substrate effect (mixed effect of strain and temperature).** (a) Peak position of 7L $E_{11}$ versus temperature on different substrates. Dots are experimental data and solid curves are fitting curves. On sapphire substrate, the samples were only cooled down to 80 K, since liquid nitrogen was used for those measurements. (b) Changing rate (extracted from the linear fittings of $E_{nn}^{N} - E_{mm}^{M}$ versus temperature) as a function of energy difference ($E_{nn}^{N} - E_{mm}^{M}$) on different substrates. Dots are experimental data and solid lines are linear fittings.